\documentclass[prl,aps,superscriptaddress,preprint,floatfix,nofootinbib]{revtex4-1}

\usepackage{epsf}
\usepackage{graphicx}
\usepackage{sidecap}

\usepackage{color}

\def \beq {\begin{equation}}
\def \eeq {\end{equation}}
\begin{document}

\title{{An experimental algorithm for identifying the topological nature of Kondo and mixed valence insulators}}



\author{M.~Neupane}
\affiliation {Joseph Henry Laboratory and Department of Physics,
Princeton University, Princeton, New Jersey 08544, USA}
\affiliation {Princeton Center for Complex Materials, Princeton University, Princeton, New Jersey 08544, USA}

\author{N.~Alidoust}\affiliation {Joseph Henry Laboratory and Department of Physics, Princeton University, Princeton, New Jersey 08544, USA}

\author{S.-Y.~Xu}
\affiliation {Joseph Henry Laboratory and Department of Physics, Princeton University, Princeton, New Jersey 08544, USA}



\author{D.J. Kim}
\affiliation {Department of Physics and Astronomy, University of California at Irvine, Irvine, CA 92697, USA}

\author{Chang~Liu}
\affiliation {Joseph Henry Laboratory and Department of Physics,
Princeton University, Princeton, New Jersey 08544, USA}

\author{I.~Belopolski}
\affiliation {Joseph Henry Laboratory and Department of Physics, Princeton University, Princeton, New Jersey 08544, USA}




\author{T.~Durakiewicz}
\affiliation {Condensed Matter and Magnet Science Group, Los Alamos National Laboratory, Los Alamos, NM 87545, USA}


\author{H.~Lin}
\affiliation {Department of Physics, Northeastern University,
Boston, Massachusetts 02115, USA}

\author{Z. Fisk}
\affiliation {Department of Physics and Astronomy, University of California at Irvine, Irvine, CA 92697, USA}

\author{A.~Bansil}
\affiliation {Department of Physics, Northeastern University,
Boston, Massachusetts 02115, USA}


\author{G.~Bian}
\affiliation {Joseph Henry Laboratory and Department of Physics, Princeton University, Princeton, New Jersey 08544, USA}

\author{M.~Z.~Hasan}
\affiliation {Joseph Henry Laboratory and Department of Physics,
Princeton University, Princeton, New Jersey 08544, USA}
\affiliation {Princeton Center for Complex Materials, Princeton University, Princeton, New Jersey 08544, USA}
\affiliation{Princeton Institute for Science and Technology of
Materials, Princeton University, Princeton, NJ 08544, USA}

\date{18 June, 2013}
\begin{abstract}

\textbf{Possible topological nature of Kondo and mixed valence insulators has been a recent topic of interest in condensed matter physics. Attention has focused on SmB$_6$, which has long been known to exhibit low temperature transport anomaly, whose origin is of independent interest. We argue that it is possible to resolve the topological nature of surface states by uniquely accessing the surface electronic structure of the low temperature anomalous transport regime through combining state-of-the-art laser- and synchrotron-based angle-resolved photoemission spectroscopy (ARPES) with or without spin resolution. A combination of low temperature and ultra-high resolution (laser) which is lacking in previous ARPES studies of this compound is the key to resolve the possible existence of topological surface state in SmB$_6$. Here we outline an experimental algorithm to systematically explore the topological versus trivial or mixed (topological and trivial SS admixture as in the first 3D TI Bi$_{1-x}$Sb$_x$) nature of the surface states in Kondo and mixed valence insulators. We conclude based on this methodology that the observed topology of the surface Fermi surface in our low temperature data considered within the level of current resolution is consistent with the theoretically predicted topological picture, suggesting a topological origin of the dominant in-gap ARPES signal in SmB$_6$.}

\end{abstract}
\date{\today}
\maketitle

Materials with strong electron correlations often exhibit exotic ground states such as the heavy fermion, Mott and Kondo insulating states, and unconventional superconductivity. Kondo insulators are mostly realized in the rare-earth based materials featuring \textit{f}-electron degrees of freedom, which behave like a correlated metal at high temperatures, with a bulk bandgap opening at low temperatures \cite{Fisk, Phil, Riseborough}. The energy gap is attributed to the hybridization of the nearly localized flat 4\textit{f} bands located near the Fermi level with the dispersive conduction band. With the advent of topological insulators \cite{Hasan, Hsieh, SCZhang, Hasan2, Xia} the compound SmB$_6$, sometimes categorized as a heavy fermion semiconductor \cite{Fisk, Phil, Riseborough}, attracted much attention due to the proposal that it may possibly host topological Kondo phase (TKI) as well \cite{Dzero,Takimoto, Dzero_2, Dai}. This compound is metallic at room temperature, but the resistivity goes up at low temperatures, consistent with its Kondo insulator (mixed-valence) nature. However, at very low temperature (T $\leq8$ K), the resistivity does not shoot up but saturates to a finite value \cite{Menth, Allen, Cooley}.
The anomalous residual conductivity present in SmB$_6$ at low temperatures is believed to be associated with the states that lie within the Kondo bandgap, indirectly evidenced in various experiments \cite{Kimura, Nanba, Nyhus, Alekseev, point_cont, Miyazaki, Denlinger, Fisk_discovery, Hall}. Here we describe a sequence for the high-resolution and low temperature angle-resolved photoemission spectroscopy (ARPES) measurements to explore the topological versus non-topological nature of surface state in Kondo and mixed valence insulators.

\vspace{0.2cm}

\textbf{Experimental sequence}

\vspace{0.1cm}
\textbf{Transport background of SmB$_6$}


SmB$_6$, a heavy fermion compound, has long been under extensive studies in transport experiments. At high and intermediate temperatures (T $ >$ 30 K), SmB$_6$ is found to behave metallic-like, but the resistivity goes up upon decreasing temperature, which suggests the occurrence of Kondo hybridization leading to a Kondo gap opened at the Fermi level. However, at very low temperatures below 7 K, the resistivity starts to saturate to a finite value (see Fig. 3c), which cannot be explained by the physical picture of a simple Kondo insulator alone. As reported in a recent transport study \cite{tunelling}, the Kondo hybridization gap is found to open at T $\sim$ 30 K. On further decreasing the temperature below 30 K, the Kondo gap increases and saturates at $\sim$ 15 meV at T  $\simeq$ 6 K. At the same temperature (T $\simeq$ 6 K), the resistivity anomaly occurs, which suggests the existence of low-lying states.

\begin{figure*}
\centering
\includegraphics[width=12cm]{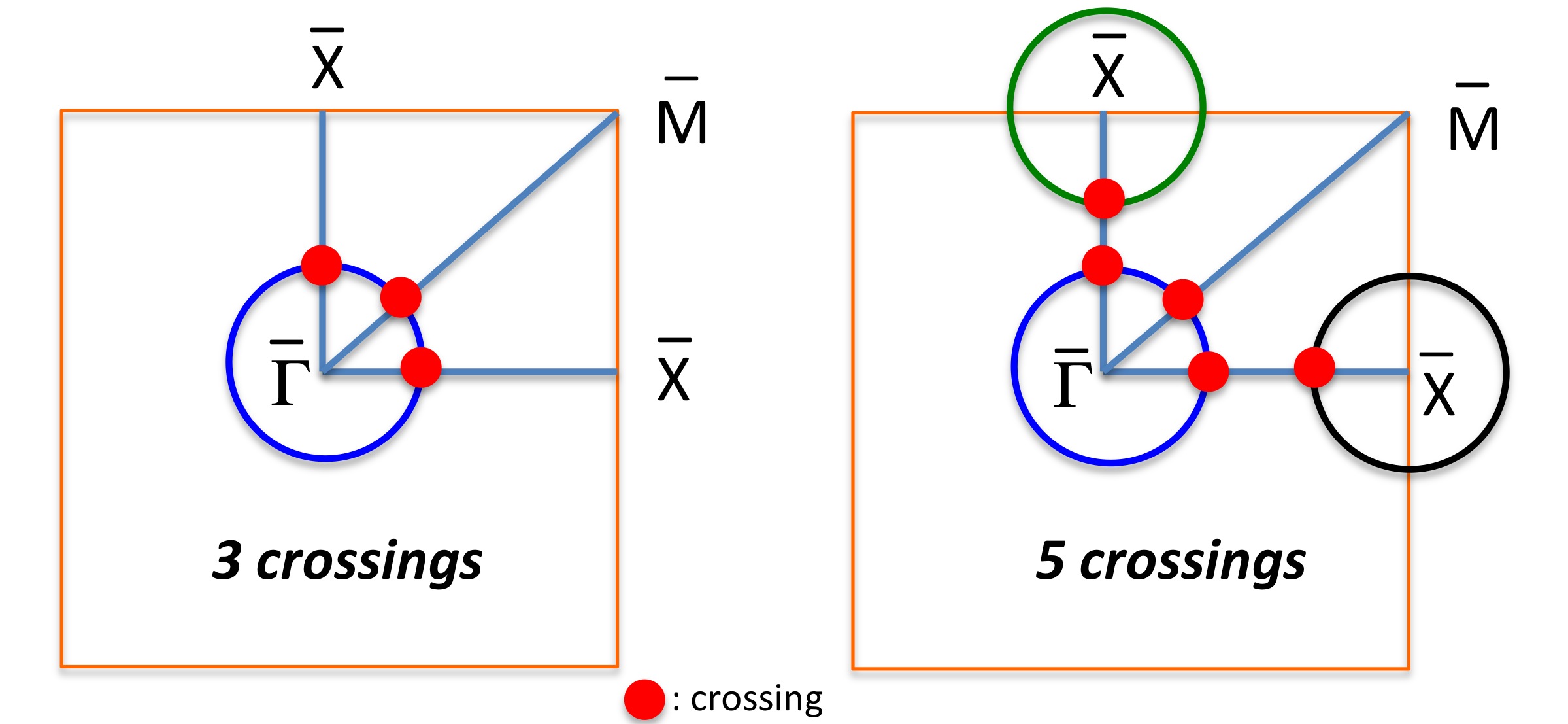}
\caption{\label{Connection} \textbf{Electronic dispersion between Kramers' degenerate points $\bar\Gamma$ and $\bar{X}$, and $\bar\Gamma$ and $\bar{M}$ in a spin-orbit system}. The number of surface state crossings at the Fermi level is odd (3 (left) and 5(right)). An odd number of crossings of \textit{non-degenerate} bands leads to the topologically nontrivial TI phase. An odd number of \textit{non-degenerate} surface Fermi pockets leads to net Berry's phase of $\pi$. Surface reconstruction or other forms of surface chemistry or dangling bond surface states must always come in pairs (even numbers) which is irrespective of the chemistry details of the surface termination.}
\end{figure*}

\vspace{0.5cm}
\textbf{Topological phase in SmB$_6$}

At temperatures T $<$ 30 K, Kondo hybridization gap gradually opens at the Fermi level. In momentum space, the hybridization gap is found to locate in the vicinity of the three $X$ points ($X_1$, $X_2$ and $X_3$) along the $\bar\Gamma-\bar{X}$ momentum space cut-direction (see Fig. 2 in ref. \cite{MN}). Very recently, it has been predicted that the Kondo insulator SmB$_6$ can host topological insulator phase, and therefore realizes a topological Kondo insulator (TKI). SmB$_6$ is actually a mixed valence compound but for the discussions focusing on topology we will continue to use the term TKI. The predicted topological surface states are consistent with the low temperature resistivity anomaly. Such a new phase of topological matter is of great interest since it enables to test the interplay between topological order and strong electron-electron correlation in reduced dimensions. According to recent theoretical studies \cite{Dai}, the bulk band-inversion occurs at the Kondo hybridization gap, which locates near the $X$ points. Since there are three $X$ points in the first BZ, there are in total three band-inversions in SmB$_6$, which leads to a nontrivial Z$_2$ topological insulator phase. This is further confirmed by the bulk band parity calculation, where the three $X$ points show negative (inverted) parity and other high symmetry points in the BZ all show positive (non-inverted) parity. At the (001) surface, these three $X$ points, ${X_1}$, ${X_2}$ and $ X_3 = {\Gamma}$, project onto the $\bar {X_1}$, $\bar {X_2}$ and $\bar {\Gamma}$, respectively (see Fig. 3a, 3b and 3d). Therefore, there exist three distinct topological surface states that enclose the $\bar {X_1}$, $\bar {X_2}$ and $\bar {\Gamma}$ points at the (001) surface (see Fig. 3a and 3b). This is indeed confirmed by the surface electronic structure calculation in \cite{Dai}.

\textbf{Conditions for the experimental demonstration of the TKI phase in SmB$_6$}

\textbf{General conditions:} It is widely accepted that ARPES and spin-resolved ARPES are the experimental tools to identify a three-dimensional topological insulator phase. Firstly, one needs to use ARPES to show that a surface Fermi surface with odd number of pockets exists.  Fixing the chemical potential at an energy level within the bulk band-gap and by traversing the momentum space from the surface Brillouin zone (BZ) center to the BZ edge at a Kramers' point, the system must feature an odd number of surface band crossing the Fermi level ($E_{\mathrm{F}}$), as shown in Fig.~\ref{Connection}. Secondly, an even number of surface state crossings will be a signature of topological triviality. In a TI, the surface states cannot come in even numbers like the Rashba surface states. Thirdly, these pockets can only be around the Kramers' points of the lattice Brillouin zone, not any other high symmetry points of the BZ. Fourthly, surface states must have non-trivial Berry's phase. Fifthly, there must be bulk-boundary correspondence and non-trivial entanglement entropy (and this is the most rigorous proof of topological order; see ref. \cite{Zhu, Wen}). For a Z$_2$ TI non-trivial entanglement is short-range.

\textbf{Temperature:} In the predicted TKI phase in SmB$_6$, the Kondo gap serves as the bulk band-gap in a regular Z$_2$ TI such as Bi$_2$Se$_3$. As mentioned above, transport experiments found that the Kondo hybridization gap gradually increases as temperature is decreased from 30 K  \cite{tunelling} and saturates at $\sim$ 15 meV when temperature reaches about 6 K. Therefore, it is critically important to work at a temperature sufficiently below 30 K so that the Kondo gap is pronounced and fully formed and the predicted topological surface states within the Kondo gap can be well-defined in energy and momentum space. Furthermore, the transport anomaly also occurs around 7 K. Therefore, to unambiguously observe the topological surface states, temperatures T $\leq$ 6 K is a necessary condition.

\begin{figure*}
\centering
\includegraphics[width=16.5cm]{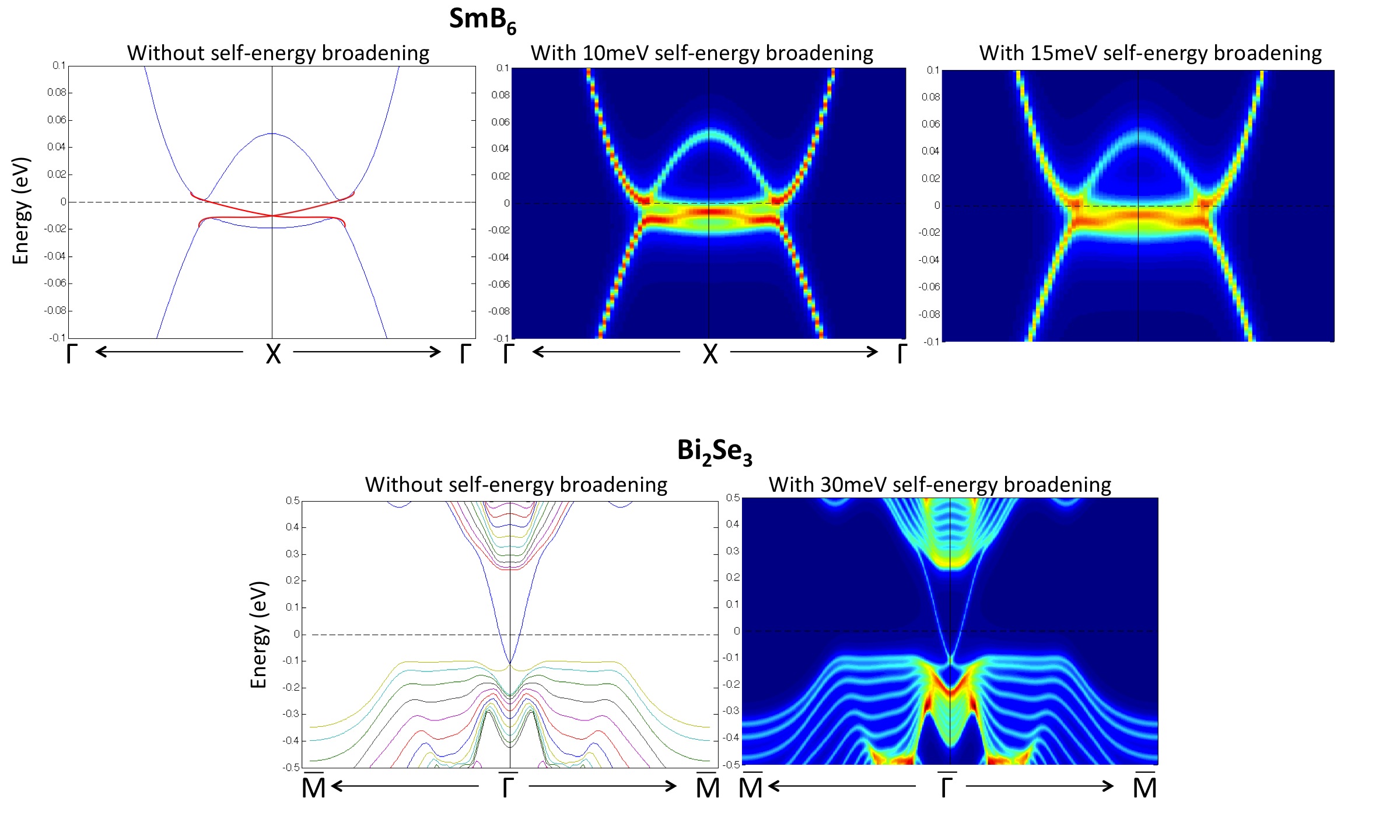}
\caption{\label{life_time}{\textbf{Bulk band structure and self-energy effects in SmB$_6$.}}
Top: First-principles calculation of the bulk band structure of SmB$_6$, showing a $\sim$ 15 meV Kondo insulating gap. Due to the fact that the Kondo gap is relatively small, it is challenging to resolve the detailed dispersion of the Dirac surface states with self-energy broadening (middle panel: 10 meV, right panel: 15 meV) is considered. Bottom: First-principles calculation on Bi$_2$Se$_3$ surface electronic structure. The bulk band gap of Bi$_2$Se$_3$ is as large as $\sim$ 300 meV. Thus, with the same self energy broadening (right panel), the surface states within the band-gap can be well resolved.}
\end{figure*}

\begin{figure*}
\centering
\includegraphics[width=12.0cm]{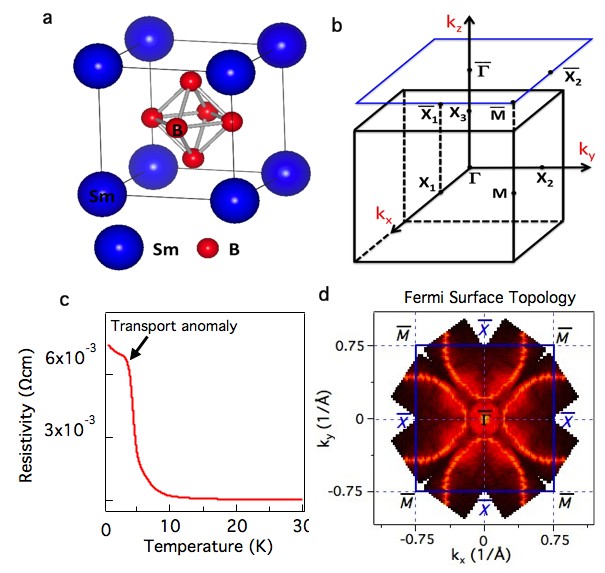}
\caption{\label{cut}{\textbf{Brillouin zone symmetry and surface Fermi surface of SmB$_6$}}. \textbf{(a)} Crystal structure of SmB$_6$. Sm ions and B$_6$ octahedron are located at the corners and the center of the cubic lattice structure. \textbf{b,} The bulk and surface Brillouin zones of SmB$_6$. High-symmetry points are marked. \textbf{c,} Resistivity-temperature profile for samples used in ARPES measurements. \textbf{(d)} Experimentally measured surface Fermi surface of SmB$_6$ \cite{MN_1}.}
\end{figure*}

\textbf{Energy resolution:} Even at temperatures T $\leq$ 6 K, the \textit{full} Kondo gap (occupied to unoccupied states) is only $\sim15$ meV. The predicted surface states lie within this Kondo gap. It is possible that Fermi level (chemical potential) lies somewhere inside the gap and it must if the bulk is insulating. Therefore, in order to resolve any dispersion of those surface states, it is critical to have an energy resolution sufficiently better than 7 meV. For example, $\Delta{E} \leq 15/2 \simeq 7$ meV would be the minimum requirement.

\textbf{Finite quasi-particle lifetime}: Assuming ARPES has a near-ideal energy resolution ($\Delta$E$  <$ 1 meV) for this problem at hand, another very important factor has to be taken into account which is the fact that the quasi-particles have finite lifetime (since the gap is small), leading to a self-energy broadening of the dispersion spectrum. For example, Fig. \ref{life_time} shows the theoretically calculated SmB$_6$ bulk $d-f$ band structure (blue lines) with a Kondo gap of $\sim$15 meV without and with a 10 or 15 meV Lorentzian broadening. It can be seen that with self-energy broadening, it becomes quite challenging to clearly track the dispersion of the surface states (red lines) at a precise qualitative level, which are within the 15 meV Kondo gap, as shown in Fig. \ref{life_time}. In contrast, similar broadening process applied on Bi$_2$Se$_3$ calculation gives a spectrum where the bands can be well-resolved, because the surface states disperse within a much larger bulk energy gap ($\sim300$ meV). Therefore, finite quasi-particle lifetime is a critical factor that makes it challenging to resolve the dispersion of surface states in SmB$_6$. Ultra-high quality SmB$_6$ samples (for example, grown by molecular beam epitaxy) with much higher transport mobility are necessary for such purposes. The above demonstration of the self-energy broadening effect in a $\sim$ 15 meV Kondo gap in SmB$_6$ also justifies our emphasis on using the partially angle (momentum) integrated ARPES spectral intensities in our earlier and present work, since it is almost impossible to resolve the E-k dispersion of the surface states in the angle (momentum) resolved ARPES dispersion maps if a 10 meV self-energy is present.

\vspace{0.5cm}

 \textbf{Spin-resolved measurements}

Recently the observation of the finite circular dichroic (CD) ARPES signal on the low energy states of SmB$_6$ is argued to be an evidence in favor of a nontrivial spin texture. However, it has been shown that CD-ARPES does not uniquely represent the spin texture profile of the topological surface states and therefore the $\pi$ Berry's phase \cite{Rader, Lanzara, Hasan_CD}. For example in ref. \cite{Rader}, CD-ARPES measurements on Bi$_2$Te$_3$ show dramatic photon energy dependence, where even the sign of the CD-ARPES signal flips as the photon energy is tuned. It is not possible to draw unique conclusion on the spin-texture of the surface states via CD-ARPES measurements. Moreover, in ref. \cite{Hasan_CD}, even the bulk bands of Bi$_2$Te$_2$Se (a TI) show strong CD signal at certain photon energies. Clearly, CD-ARPES cannot be used to provide a unique measure of the surface spin texture or Berry's phase.

Spin-resolved ARPES is currently not feasible to resolve the in-gap states in SmB$_6$ due to the limitations on energy/momentum resolution and temperature limit of such an ARPES system. Typically the energy resolution of SR-ARPES at synchrotrons is about 30 to 50 meV, which is much larger than the 5 meV Kondo scale [see \cite{reso_1}]. With laser sources, the energy resolution is about 15 meV, still larger than the 5 meV (see \cite{reso_2}). The lowest accessible temperature with the best spin-resolution is about 20 K, which is higher than the transport anomaly (8 K). With the latest technology at hand it is currently not possible to prove points four and five of general conditions for SmB$_6$ at hand (noted in the previous section).





The low-temperature ($\sim$ 6 K) and high resolution ($\sim$ 4 meV and 0.1$^\circ$) ARPES measurements on SmB$_6$ have been reported in our previous work \cite{MN}. These low temperature and high energy resolution measurements, for the first time, provide access to study the resistivity anomaly ($\leq$ 8 K) as well as the predicted in-gap surface states (within the 15-20 meV (theoretical value) Kondo gap below 6 K). Firstly, at very low temperature (T $=$ 6 K, corresponding to the transport anomaly), we observe a pronounced ARPES in-gap spectral intensity within a $\sim$ 4 meV energy window that lies inside the $\sim$ 17 meV Kondo insulating gap (see Fig 2 in \cite{MN}), which experimentally shows the existence of in-gap states. Secondly, the dominant signals from the in-gap states are found to be suppressed upon raising the temperature and disappear before 30 K, which correspond to the Kondo hybridization onset temperature at which the Kondo gap closes. Such in-gap states are further observed to be robust against thermal cycling (see 6 K thermal recycling in Fig 2 of \cite{MN}), consistent with their topologically protected nature predicted in theory. Thirdly, ARPES "Fermi surface mapping" including a window to cover the energy of the in-gap states shows three distinct Fermi pockets that enclose the $\bar {X_1}$, $\bar {X_2}$ and $\bar {\Gamma}$ points respectively (see Fig. 3d), which is consistent with the predicted Fermi surface topology for the topological surface states of the Kondo phase in SmB$_6$.

Furthermore, our k-space map covering the in-gap states (Fig. 3d) shows distinct pockets that enclose three (not four) Kramers' points of the surface Brillouin zone, which are remarkably consistent with the theoretically predicted topological Fermi surface in the TKI groundstate. Previous and present bulk band calculations have reconfirmed the strong impact of spin-orbit coupling in the bulk bands in this inversion symmetric material, suggestive of the surface states (trivial or non-trivial or their admixure) on this compound being non-degenerate. This allows us to compare the surface states with theory that winds around the Kramers' points which is directly tied to the question of topology.
In the presence of bulk spin-orbit coupling and band inversion these surface states (assuming same type or handedness of their spin-textures) must carry 3$\pi$ Berry's phase which is equivalent to $\pi$ (3$\pi$ Mod 2$\pi$). According to ref. \cite{FuKaneMele} the most important criterion for a topological insulator is that odd number (1, 3, 5, ...) of surface Fermi pockets of non-degenerate bands exist in a strong spin-orbit system with bulk band inversion. These theoretical criteria directly imply a \textit{net} surface Berry's phase of $\pi$ \cite{FuKaneMele}. Another independent way of proving a total 3$\pi$ Berry's phase in this system would be to carry out a direct spin-ARPES measurement as demonstrated previously in Bi-based compounds in many of our earlier works (see reviews \cite{Hasan, SCZhang, Hsieh}). It is important to stress that the \textit{dominant} in-gap states are found to be robust and reproducible against thermal cycling of the sample in and out of the Kondo regime only (see Fig. 2 of \cite{MN}), excluding the possibility of unrelated trivial surface states (as dominant signals) due to dangling bonds or polarity-driven states which can robustly appear at much high temperatures (ignoring the Kondo physics) due to surface chemistry. Within the level of our resolution we do see some weak signal like that but these weak (non-dominant) signal is not robust. Polarity driven (non-topological) surface states must also come in even number of pockets at any high symmetry point of the Brillouin zone unrelated to the Kramers' point winding behavior but that can add the the spectral weight. Like in the first 3D topological insulator Bi$_{1-x}$Sb$_x$ system \cite{Hsieh}, SmB$_6$ may also host topological and trivial surface states admixtures but even higher energy resolution would be necessary to decisively resolve their character. Coexistence of topological and trivial surface states does not make the system itself (SmB$_6$) topologically trivial. It is important to note that if the pockets around the $\bar{X}$ points were to originate from the bulk band, SmB$_6$ would be a bulk metal (in carefully grown samples this is not the case). Transport measurements on our samples do clearly show the bulk insulating behavior \cite{Hall}, supporting the absence of the bulk pockets around the $\bar{X}$ points.


The observed Fermi surface topology of the in-gap states, as well as their temperature dependence across the transport anomaly and Kondo hybridization temperature, collectively not only provide a unique insight illuminating the nature of the residual conductivity anomaly but also serve as a strong evidence in support of a topological Kondo insulator phase in SmB$_6$. We note that the studies of dominant in-gap states within the 4 meV energy window within the Kondo gap using high resolution ARPES are partly based on the angle (momentum) integrated ARPES data. The finer details of the energy and momentum dispersion of the dominant in-gap surface states cannot be obtained in the angle (momentum) resolved mode due to intrinsic self-energy/resolution broadening (even given the SmB$_6$ samples with high mobility) in a 15 meV Kondo gap, as discussed above. Furthermore, one can speculate that the lack of k$_\mathrm{z}$ dispersion in the bands lying above $E_B =$10 meV is due to some band-bending effect of the bulk bands that must occur in an intrinsic bulk insulator/semiconductor. A bent-bulk band near a surface should naturally exhibit weak k$_z$ dispersion since the effect is confined near the surface only and surface bands must also emerge from the bent bulk bands in all topological band insulators. In order to isolate the in-gap states from the bulk band tails that have higher cross-section at synchrotron photon energies, it is necessary to have energy resolution better than half of the Kondo gap scale, which is about 7 meV or smaller in SmB$_6$. Some details regarding the issues with band bending in TIs are discussed our earlier works \cite{HsiehNat, Andrew}.

\vspace{0.5cm}

\textbf{In-gap states}

\begin{figure*}
\centering
\includegraphics[width=16.0cm]{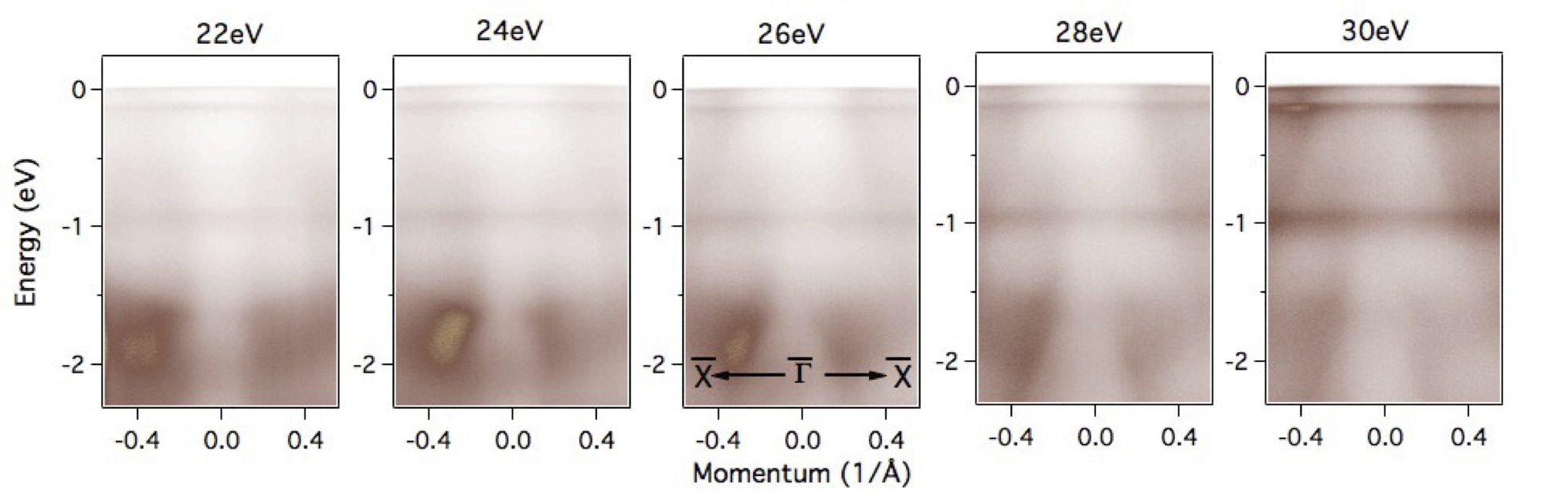}
\caption{\label{photon}{ARPES dispersion map at the $\bar{X}-\bar\Gamma-\bar{X}$ momentum space cut}. These  ARPES dispersion spectra are measured at SSRL beamline 5-4 with different photon energies at temperature of 10 K over the wide binding energy range.}
\end{figure*}

The temperature dependent behavior of states near the Fermi level is observed at both the $\bar\Gamma$ and the $\bar{X}$ points (see Fig. 2 in \cite{MN}). With a 7 eV laser-based photoemission system, photoionization cross section for the $f$ orbitals is weak, so the measured states should significantly correspond to the partial $d$ orbital character of the hybridized band if the hybridization exists.
Based on the temperature dependent electronic structure presented in Fig. 2 of \cite{MN}, we point out two relevant temperatures which are around 15 K and 30 K. We note that these in-gap states are robust below 15 K, appear to be weak above 15 K and finally vanish above 30 K (Figs. 2c and d in ref. \cite{MN}). The robustness of these states below 15 K is in correspondence with the 2D conductivity channels in SmB$_6$. It is important to note that the in-gap states at $\bar\Gamma$ can only be observed with high-resolution measurements ($\sim$ 4 meV and 0.1$^{\circ}$) performed at low temperature ($<$ 8 K). This reveals the importance of high energy-momentum resolution for the ARPES measurements in order to gain useful insights ( Figure in 3 in ref. \cite{MN} versus Figure \ref{photon}).

\begin{figure*}
\centering
\includegraphics[width=16.5cm]{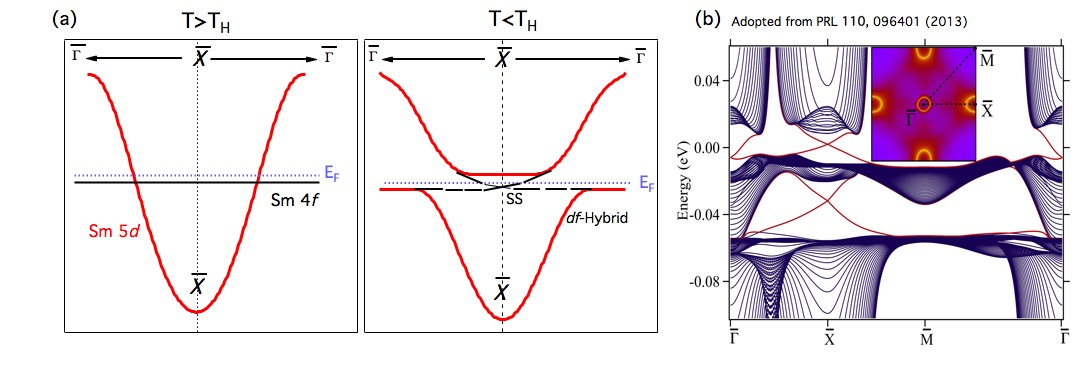}
\caption{\label{in_gap}{\textbf{Surface states within the hybridization gap.}}
(\textbf{a}) Schematic view of the hybridization above and below the hybridization temperature.
(\textbf{b}) Calculation of surface electronic structure taken from ref. \cite{Dai}.}
\end{figure*}

\begin{figure*}
\centering
\includegraphics[width=16.5cm]{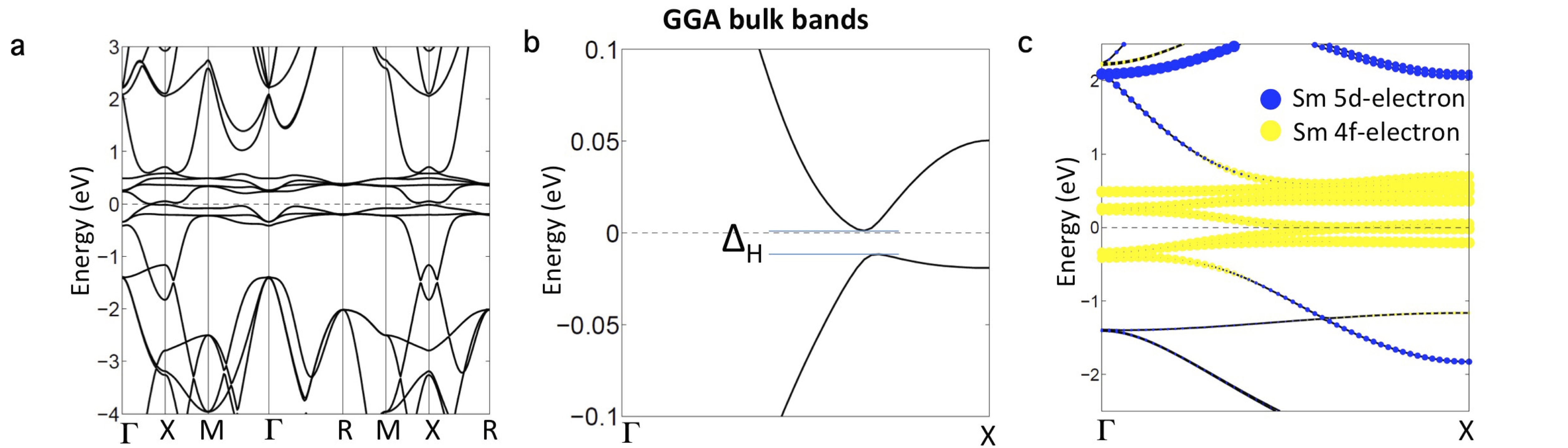}
\caption{\label{cal}{\textbf{Bulk band structure of SmB$_6$.}}
(\textbf{a}) The calculated band structure of SmB$_6$ by GGA. (\textbf{b}) Zoom in of (a) near Fermi level along $\Gamma-{{X}}$. The bandgap of about 15 meV was obtained in the calculation.
(\textbf{c}) Orbital decomposed band structure near the Fermi level along $\Gamma-{{X}}$. The size of the blue and yellow spheres is proportional to the weight of Sm 5$d$ and 4$f$ orbitals, respectively.}
\end{figure*}

\vspace{0.5cm}

 \textbf{Bulk band calculations of SmB$_6$}

In order to further cross check the bulk band dispersion, we performed first-principles bulk band calculations based on the generalized gradient approximation (GGA) \cite{GGA} using the projector augmented-wave method \cite{PAW, PAW_1} as implemented in the VASP package \cite{VASP, VASP_1}. The experimental crystallographic structure \cite{expt} was used for our calculations. The spin-orbit coupling was included self-consistently in the electronic structure calculations with a 12$\times$12$\times$12 Monkhorst-Pack $k-$mesh. Fig. \ref{cal} shows the calculated bulk band structure of SmB$_6$, which agrees with previous calculations \cite{Dzero, Dzero_2, Takimoto, Kai, Dai}. With fine tuned parameters, our GGA calculations give an insulating ground state with a small energy gap of 15 meV (Fig.  \ref{cal}b). From the orbital-decomposed band structure (Fig. \ref{cal}c), we find that the flat Sm 4$f$ bands are located around E$_F$ from -0.5 eV. An itinerant Sm 5$d$ band with larger dispersion hybridizes with the 4$f$ bands, forming an anti-crossing band shape.
Our GGA calculations qualitatively agree with the observed bulk band dispersion.

In conclusion, we argue that laser- and synchrotron-based systematic ARPES measurements can be used to resolve the issue of topological versus trivial (or mixed as in Bi-Sb system) nature of the surface states. With high resolution and low temperature laser- and SR-ARPES data, we observe that firstly, at low temperatures T $\sim$ 6 K (corresponding to the transport anomaly), pronounced ARPES spectral intensity exists within the Kondo gap ($\sim$ 4 meV) energy window lying inside the $\sim$ 17 meV insulating gap, which experimentally shows the existence of in-gap states. Secondly, the dominant part of the in-gap states are found to be suppressed while raising the temperature and disappear before reaching T $=$ 30 K, which correspond to the Kondo hybridization onset temperature at which the Kondo gap closes. Such in-gap states are further observed to be robust against thermal cycling, consistent with their protected/robust nature. Thirdly, ARPES Fermi surface or iso-energy contour measurements at energies that correspond to the in-gap states show three (odd as opposed to even) distinct Fermi pockets within our experimental resolution that enclose the $\bar {X_1}$, $\bar {X_2}$ and $\bar {\Gamma}$ points only, respectively. Fourthly, the observed Fermi surface topology of the dominant in-gap states, as well as their temperature dependence across the transport anomaly and Kondo hybridization temperatures, collectively not only provide a unique insight illuminating the nature of the residual conductivity anomaly but also serve as a strong supporting evidence for the existence of a topological Kondo insulator phase in SmB$_6$ within the level of current state-of-the-art experimental resolution.


\end{document}